\documentclass[12pt]{article}
\usepackage{amsmath,comment}
\usepackage{graphicx}
\graphicspath{plot/}
\usepackage{enumerate}
\usepackage{natbib}
\usepackage{url} 

\usepackage{amsmath}
\usepackage{amsthm}
\usepackage{amssymb}
\usepackage{amsfonts}
\usepackage{array}
\usepackage{relsize}
\usepackage[ruled,vlined]{algorithm2e}
\SetKwInput{kwInit}{Init}
\usepackage{tabularx} 

\usepackage{caption}
\usepackage{subcaption}  
\usepackage{multirow}
\usepackage{tabularx}
\usepackage{algorithm2e,setspace}
\usepackage[noend]{algpseudocode}
\usepackage{float}
\usepackage{booktabs}
\usepackage{xcolor}
\usepackage{cancel}

\usepackage[plainpages=false,pdfborder={0 0 0}]{hyperref}

\newtheorem{definition}{\textsc{Definition}}

\newcommand{\blind}{1}

\begin{document}

\def\spacingset#1{\renewcommand{\baselinestretch}%
{#1}\small\normalsize} \spacingset{1}


\if1\blind
{
  \title{\bf Spatial homogeneity learning for spatially correlated functional data with application to COVID-19 Growth rate curves}
  \author{Tianyu Pan, Weining Shen, \hspace{.2cm}\\
    Department of Statistics, University of California, Irvine\\
    and \\
    Guanyu Hu\thanks{Email: \url{guanyu.hu@missouri.edu}} \\
    University of Missouri - Columbia, Columbia, MO, 65211}
  \maketitle
} \fi

\if0\blind 
54

{
  \bigskip
  \bigskip
  \bigskip
  \begin{center}
    {\LARGE\bf Title}
\end{center}
  \medskip
} \fi

\bigskip
\begin{abstract}
We study the spatial heterogeneity effect on regional COVID-19 pandemic timing and severity by analyzing the COVID-19 growth rate curves in the United States. We propose a geographically detailed functional data grouping method equipped with a functional conditional autoregressive (CAR) prior to fully capture the spatial
correlation in the pandemic curves. The spatial homogeneity pattern can then be detected by a geographically weighted Chinese restaurant process prior which allows
both locally spatially contiguous groups and globally discontiguous groups. We design an efficient Markov chain Monte Carlo (MCMC) algorithm to simultaneously infer the posterior distributions of the number of groups and the grouping configuration
of spatial functional data. The superior numerical performance of the proposed method over competing methods is demonstrated using simulated studies and an application to COVID-19  state-level and county-level data study in the United States. \\

\noindent%
\end{abstract}

\vfill

\newpage
\spacingset{1.5} 

\section{Introduction}\label{sec:intro}

The ongoing pandemic of novel coronavirus disease (COVID-19) has become a worldwide public health issue  since December 2019, and it has landed a detrimental effect on every aspect of human lives.   
There is an emerging literature in statistics studying COVID-19 data, with the majority  \citep{wu2020nowcasting,chen2020time,read2020novel,tang2020estimation,sun2020discussion,hu2020heterogeneity,yang2020time} focusing on the analysis and prediction of the daily confirmed, recovered, and reproduced cases based on Susceptible-Infectious-Recovered (SIR) model and its variations \citep{kermack1932contributions,kermack1933contributions}. However, appropriate statistical models are still largely needed towards a deeper understanding of the COVID-19 epidemic curves, and, more importantly, their dynamic changes over different geographic regions (e.g., states) in the United States. Epidemic curve serves as an extremely useful visualization and data exploration tool in epidemiology as it provides a direct measurement of disease progression over time (e.g., size, pattern of spread, and time trend). Moreover, understanding the similarity/disparity in epidemic curves across different regions may shed light on studying the effect of government social/economic policies on the disease progression, exploring the spatial spread pattern of the disease, and ultimately assisting future pandemic forecast and the real-time public health decision making. Several studies have suggested that the spatial heterogeneity can produce a dramatic difference in social exposures to COVID-19, and stress local healthcare systems differently in timing and severity \citep{thomas2020spatial}. This is also observed in our preliminary data analysis. For example, in Figure \ref{fig:one}, there is a clear spatial dependence pattern among contagious states, e.g., similar epidemic curves between New York and New Jersey. On the other hand, several states including California, Texas, and Florida, despite  being geographically far apart, share a similar pattern in their epidemic curves (see, Figure \ref{fig:four}), which may be related to their similar reopening policies.




The main goal of this paper is to develop a new clustering method and further the understanding of the spatial heterogeneity effect of COVID-19 epidemic curves. A desired clustering approach should be capable of taking account for the potential spatial heterogeneity and revealing interpretable latent patterns in the epidemic curves at different levels (e.g., states and counties). In statistics, it is natural to represent the epidemic curve in the form of the {\it spatially correlated functional data}, where the sampling unit can be viewed as a function over a continuous range of time collected at a geographic region. In the literature, most existing clustering approaches for functional/longitudinal data analysis are either distance-based \citep{ferraty2006nonparametric,cuesta2007impartial,genolini2010kml,hu2020bayesian} or they ignore the correlation of the functional data among different locations \citep{srivastava2020understanding}. To quote Tobler's first law of geography \citep{tobler1970computer}: ``everything is related to everything else, but near things are more related than distant things''. For spatial data, observations from nearby locations are expected to have a stronger correlation than those from distant locations; and it is hence important to incorporate such constraint when conducting spatial clustering detection \citep{knorr2000bayesian,lee2017cluster,li2019spatial,yun2020detection}. Moreover, for epidemiology data analysis, only considering spatially contiguous clusters is not good enough since there are other demographic factors (e.g., GDP, population, temperature, government policy) that may have a significant effect on clustering configurations of the epidemic curves. For example, California has a very similar COVID-19 growth pattern with that of the New York at the beginning of the outbreak despite these two states are geographically far apart. One possible explanation is that both states have a similar population density and serve as the hub for global traveling. For those reasons, it is challenging-yet-necessary to take both spatially contiguous clusters and spatially discontiguous clusters into consideration in our analysis. 
Another challenge in clustering analysis is to determine the number of clusters. The most common solution is to pre-specify the number of clusters based on certain empirical criteria \citep{jacques2014model,liang2020modeling}. Despite its computational convenience, this strategy does not take the uncertainty associated with cluster number selection into account when conducting the inference for the final clustering results.  


To overcome the aforementioned challenges, in this paper, we propose a Bayesian nonparametric method for clustering spatially correlated functional data. The proposed method provides a useful model-based clustering solution for epidemic curve study that is able to recover spatially contiguous and discontiguous clusters simultaneously without pre-specifying the number of clusters. The key novelty lies in representing the latent clustering structure in the functional data by a generalized Chinese Restaurant Process (CRP) and incorporating geographic information when sampling from the CRP. In addition, we propose an efficient Markov chain Monte Carlo (MCMC) algorithm that bypasses the need of implementing computationally expensive reversible jump MCMC/sampler allocations; and the full inference for both the number of clusters and the clustering membership can be conveniently conducted under the Bayesian framework. The proposed method is applied to study the U.S. COVID-19 data and is shown to be helpful in revealing meaningful spatial dynamic patterns of the COVID-19 progression.


The rest of this paper is organized as follows. We discuss the motivating COVID-19 data example in Section \ref{sec:data}. We introduce a model for spatially correlated functional data in Section \ref{sec:spatial_correlated_function}, followed by a discussion of nonparametric Bayesian clustering method in Section \ref{sec:gwCRP}, and the new spatial homogeneity learning method for spatially correlated functional data in Section \ref{sec:hier_model}. In Section \ref{sec:bayesian_inference}, we provide details about the Bayesian inference, including the sampling algorithm, the model selection criteria for tuning parameter, and post-MCMC inference. Simulation studies and the U.S. COVID-19 data analysis are presented in Section \ref{sec:simu} and \ref{sec:real_data}, respectively. We conclude with a discussion in Section \ref{sec:discussion}.

\section{Motivating Example}\label{sec:data}

\begin{figure}[htp]
\begin{center}
\centerline{\includegraphics[width=1\textwidth]{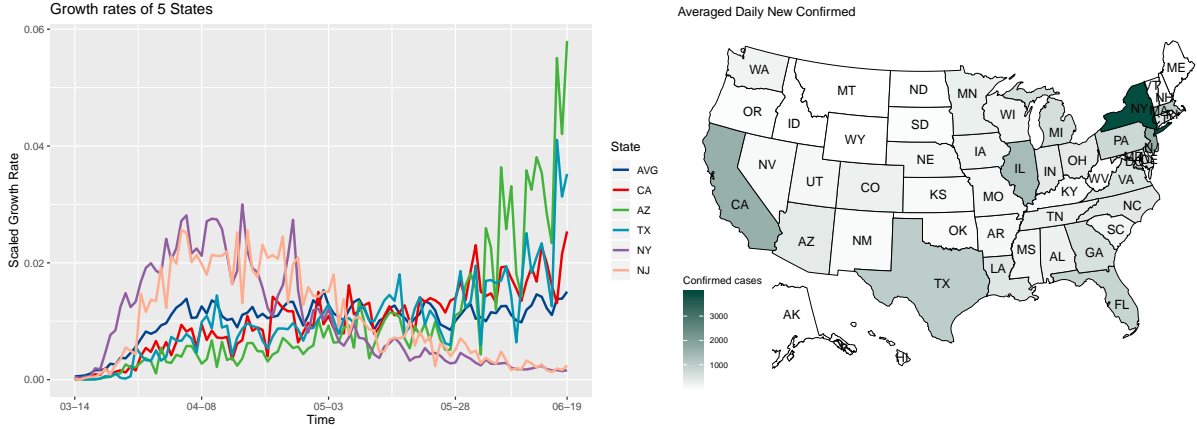}}
\end{center}
\caption{\label{fig:one}Left: Scaled growth rate curve of California, Arizona, Texas, New York, and New Jersey; Right: Averaged daily new confirmed cases by 51 States.}
\end{figure}

We consider the data collected by the COVID tracking project \url{https://covidtracking.com}.
State-level COVID-19 confirmed cases are recorded on a daily basis for the 50
states plus Washington, DC. For simplicity, we refer to them as ``51 states'' for the rest of this paper. We focus on the time frame starting from March-13th, the date when President Trump declared the state of emergency, to June-19th (a total of 99 days). To obtain the epidemic curves, we follow \citet{srivastava2020understanding} to preprocess the data as in the following steps,
\begin{enumerate}
    \item  Denote the cumulative confirmed cases for state $i$ on day $t$ by $f_i(t)$. Then the newly confirmed case on day $t$ is  defined as $s_i(t)\equiv f_i(t)-f_{i}(t-1)$.
    \item  Calculate the scaled growth rate on day $t$, which is defined as $Y_{i}(t)=\frac{s_i(t)}{\sum_{u=2}^{99} s_i(u)}$ for $t=2,3,\cdots,99$, and then rescale the time points $t=2,3,\cdots 99$ to a unit interval with $t'=j/97$ for $j=0,1,\cdots,97$ representing 98 time points in the original scale. 
    For the rest of this paper, $Y_i(t)$ is also named as \textit{scaled growth rate curve}. 
    \item In case of invalid entries, e.g., $s_i(t)<0$, we trace back to the day when the miscount takes place and remove the falsely-counted positive cases from $s_i(t)$. 
\end{enumerate}
As a demonstration, we plot the scaled growth rate curve for five chosen states and the nationwide average at the left side of Figure \ref{fig:one}. One nice feature of the scaled growth rate curve is that the population size effect is removed when studying the dynamic changes of the curve. For example, the curves of New York and New Jersey are similar despite their significant difference in their total number of confirmed cases as shown in the right side of Figure \ref{fig:one}, which can be explained by the fact that the New York population is about 2.7 times that of New Jersey.


From Figure \ref{fig:one}, we first observe a significant difference between the overall average trend (``AVG'') and the curves for selected five states, which highlights the necessity of considering spatial heterogeneity. Secondly, geographically contiguous states tend to have a similar pattern in their growth rater curves, e.g., curves from New York and New Jersey display a common `increasing-then-decreasing' pattern, and the curves from California and Arizona display a steadily increasing trend. Thirdly, states that are geographically apart (e.g., California and Texas) may still exhibit a common pattern in their curves, which may be related to other demographic factors in common such as weather and reopening policy. To synthesize these preliminary finding into a formal statistical investigation while accounting for the spatial heterogeneity in the growth rate curves, a new model-based clustering approach is needed. 


\section{Method}\label{sec:method}
\subsection{Spatially Correlated Functional Model}\label{sec:spatial_correlated_function}
For state $i$, we propose to model its scaled growth rate at time $t$, denoted by $Y_i(t)$, as
$$Y_i(t)=\mu_{c_i}(t)+\epsilon_i(t),$$
where $c_i$ represents the cluster, in which the state $i$ is allocated, the mean function for cluster $c_i$ is denoted by $\mu_{c_i}(t)$, and the residual is assumed to satisfy $\epsilon_{i}(t)\sim \mathcal{N}(0,\sigma_{c_i}^2)$ independently for every $t$. The mean function can be further expanded as
$$\mu_{c_i}(t)=\sum_{j=1}^\infty \beta_{c_i,j}\xi_j(t),$$
where we assume that the time domain is $[0,1]$ without loss of generality, and $\xi_1,\xi_2,\cdots$ are the orthonormal basis functions on $[0,1]$ satisfying
\begin{equation}
\int_0^1\xi_i(t)\xi_j(t)dt= \left\{
             \begin{array}{lr}
             0, & i\neq j \\
             1, & i=j
             \end{array}
\right..
\end{equation}
To take account for the spatial correlation between different states, we consider a conditional autoregressive prior \citep[CAR;][]{de2012bayesian,zhang2016functional} on the standardized residuals, denoted by $\epsilon_i^*(t)\equiv\frac{\epsilon_i(t)}{\sigma_{c_i}} $ as follows, 
$$[\epsilon_1^*(t),\epsilon_2^*(t),\ldots,\epsilon_n^*(t)]^T\sim \mathcal{N}(0,I-\phi A),\ \phi\sim \text{Unif}(\ell_A,u_A),$$
where the matrix $A$ is the adjacency matrix for $n$ states, and $\ell_A$ and $u_A$ refer to the reciprocal of the minimum (always taking a negative value) and maximum eigenvalue (always taking a positive value) of $A$, respectively. In practice, we truncate the infinite series of mean function $\mu_{c_i}(t)$, and choose a sufficiently large number $p$ for estimation, which results in
\begin{align}\label{eq1}
Y_i(t)=\sum_{j=1}^p \beta_{c_i,j}\xi_j(t)+\epsilon_i(t).
\end{align}
This model hence becomes a Bayesian regression model specified with a CAR covariance structure; and it is natural to assign a multivariate normal prior on $\{\beta_{c_i,j}\}_{j=1}^p$, and a conjugate inverse Gamma prior on $\sigma^2_{c_i,j}$, namely,
$$[\beta_{c_i,1},\cdots,\beta_{c_i,p}]^T\mid \sigma^2_{c_i,j}\sim \mathcal{N}(\mu_0,\sigma^2_{c_i,j}\Lambda^{-1}_0), 1/\sigma^2_{c_i,j}\sim \text{Gamma}(\nu_0,\nu_0s_0^2/2).$$
In the numerical analysis, we use the orthonormalized B-spline basis as the basis $\xi_j$'s, and the number of orthonormal basis $p$ is chosen to be $1+p'$, where $p'$ is the total number of the eigenfunctions selected by the functional principal component analysis (FPCA). More details about the choices for the hyper-parameters $\Lambda_0$, $s_0$ and $\nu_0$  will be provided in Section \ref{sec:hier_model}. 


\subsection{Geographically Weighted Chinese Restaurant Process}\label{sec:gwCRP}

In order to cluster the scaled growth rate curves, we consider a flexible nonparametric Bayesian approach, where a nonparametric prior is placed on the mixture probability and the inference is conducted simultaneously on the number of clusters, denoted by $K$, and the corresponding cluster configurations (e.g., membership and cluster-wise parameters). We start with a brief review of Dirichlet Process \citep[DP;][]{ferguson1973bayesian} and its connection to Chinese Restaurant Process \citep[CRP;][]{pitman1995exchangeable,neal2000markov} since both concepts serve as the building blocks of our proposed method. Consider a Dirichlet process DP$(\alpha,G_0)$, where $\alpha > 0$ is the concentration parameter that controls the precision of DP and $G_0$ is a probability measure that can be understood as the mean of the DP. Due to the discrete nature of a Dirichlet Process, one can always obtain a partition $\mathcal{C}$ of $[n]\equiv\{1,2,\cdots,n\}$ that corresponds to a solution to the clustering problem. It has been shown that the probability mass function \citep{antoniak1974mixtures,green2001modelling} for partition $\mathcal{C}$ is 
$$p(\mathcal{C})=\frac{\alpha^d\Pi_{i=1}^K (|c_i|-1)!}{\Gamma(\alpha+n)/\Gamma(\alpha)},$$
where $|c_i|$ is the size of cluster $c_i$ and $K$ is the number of clusters in $\mathcal{C}$. The Chinese Restaurant Process, also known as P\'{o}lya Urn Scheme, makes sampling partitions from this probability mass function feasiable by considering the following proposal, 
\begin{equation}\label{eq2}
P(n\in c\mid[n-1])\propto \left\{
             \begin{array}{lr}
             |c|, & c\in\mathcal{C}_{n-1} \\
             \alpha, & c\notin \mathcal{C}_{n-1}
             \end{array}
\right.,
\end{equation}
where $\mathcal{C}_{n-1}$ denotes a partition of $[n-1]$. One interpretation for this process is that, a new customer $n$ that entered the restaurant would either sit at one of the existing tables $c$ with a probability proportional to the number of customers currently sitting at this table, i.e., $|c|$, or start a new table with a probability proportional to $\alpha$.

In our problem, we treat the growth curve for each state as a customer in CRP, and let $\theta_c \in \mathbb{R}^{p} \times \mathbb{R}^+$ denote the collection of parameters including basis coefficients $\beta_c = (\beta_{c,1},\ldots,\beta_{c,p})^T$ and residual standard deviation $\sigma_c$ for cluster $c$ as defined in \eqref{eq1}. Note that the sampling scheme in \eqref{eq2} does not incorporate the useful spatial information. Inspired by the geographically weighted Dirichlet Process (gwDP) proposed by \citet{geng2020bayesian} for survival model, we consider a gwDP prior for the functional data clustering purpose, whose predictive distribution is given by the following definition. 

\begin{definition}
\label{def:gwDP}
Let $G_0$ be a continuous probability measure on $\mathbb{R}^{p} \times \mathbb{R}^+$. We define the predictive distribution of $\theta_{c_n}$ given $\theta_{c_1},\cdots,\theta_{c_{n-1}}$ as 
\[\Pi(\theta_{c_n}\mid\{\theta_{c_i}\}_{i=1}^{n-1})\propto\sum_{c\in\mathcal{C}_{n-1}}\sum_{j\in c}w_{n,j}\delta_{\theta_c}(\theta_{c_n})+\alpha G_0(\theta_{c_n}),\]
\end{definition}
where $c_n$ represents the cluster that state $n$ is allocated,  $\theta_c$ is the parameter shared within the cluster $c$, $\mathcal{C}_{n-1}$ is the partition for $[n-1]$, $\delta(\cdot)$ is the indicator function, and $w_{i,j}\in [0,1]$ are elements in the weighted symmetric matrix $W\equiv (w_{i,j})_{n\times n}$ that specifies the spatial relationship between state $i$ and $j$. Similarly with \eqref{eq2}, we can further define a geographically weighted Chinese Restaurant Process (gwCRP) for gwDP by considering 
\begin{equation}
P(n\in c\mid[n-1])\propto \left\{
             \begin{array}{lr}
             \sum_{j\in c}w_{n,j}, & c\in\mathcal{C}_{n-1} \\
             \alpha, & c\notin \mathcal{C}_{n-1}
             \end{array}
\right..
\end{equation}
It is clear that the proposed gwDP and gwCRP are generalizations of the classical DP and CRP, i.e., if all the weights in the weight matrix $W$ equal to $1$, then we obtain CRP as a special case. In our problem, to take account for the geographical relationship, we adopt the choice of the geometric weights $w_{i,j}$ as follows, 
\begin{equation}\label{eq5}
w_{i,j}= \left\{
             \begin{array}{lr}
             1, & \text{if}\ d_{i,j}\leq 1\\
             \exp\{-d_{i,j}h\}, & \text{if}\ d_{i,j}>1
             \end{array}
\right., i,j=1,\ldots,n,
\end{equation}
where $h \geq 0$ is a tuning parameter  representing the strengh of association between the distance $d_{ij}$ and the spatial correlation in the model for states $i$ and $j$.  This weight choice has also been used in survival analysis of spatial data \citep{xue2019geographically}. Other choices of the weight functions can also be adopted. Note that as $h=0$, the gwCRP degenerates to the conventional CRP. If $h\to\infty$, the resulting gwCRP will only concern the adjacent states. The choice of $h$ is hence important and will be discussed in Section~\ref{sec:model_select}.  In our work, $d_{i,j}$ is calculated by implementing the Dijkstra Algorithm on the adjacency matrix.

\subsection{Hierarchical Model and prior specification}\label{sec:hier_model}
Now we are ready to present the full hierarchical model and discuss the choice of the prior distribution in this section. Let $Y_i=(Y_i(1),\ldots,Y_i(T))^T$ be the collection of observed functional data for state $i$ over $T$ time points, $\xi_j=(\xi_j(1),\ldots,\xi_j(T))^T$ be the collection of basis functions, and  $\beta_{c_i}=(\beta_{c_i,1},\ldots,\beta_{c_i,p})^T$ be the basis expansion coefficients. For state $i$ 
that belongs to cluster $c_i$, let $\theta_i = \{\beta_i,\sigma^2_i\}$ and note that $\theta_i$ takes the same value for every $i$ that belongs to the same cluster. Our proposed model can then be presented in the following hierarchical structure,
\begin{equation}
    \begin{split}
     & \text{\text{vec}}\left([\frac{Y_i-[\xi_1,\cdots,\xi_p]\beta_i}{\sigma_i}]_{i\in[n],T\times n}\right)\mid \{\theta_i\}_{i=1}^n\sim \mathcal{N}(0,(I-\phi A)_{n\times n}^{-1}\otimes I_{T\times T}),\\
&\{\theta_i\}_{i=1}^{n}\mid G\sim G,\\
&G\mid\alpha,G_0,W(h)\sim \text{gwDP}(\alpha,G_0,W(h)),\\
&d G_0\equiv \pi(\beta,\sigma^2)d\beta d\sigma^2\\
&\beta\mid\sigma^2\sim \mathcal{N}(\mu_0,\sigma^2\Lambda_0^{-1}),\\
&1/\sigma^2\sim \text{Gamma}(\text{shape}=\nu_0/2,\text{rate}=\nu_0\times s_0^2/2),\\
&\phi \sim \text{Unif}(\ell_{A},u_{A}),\\  
    \end{split}
\end{equation}
where $[\frac{Y_i-[\xi_1,\cdots,\xi_p]\beta_i}{\sigma_i}]_{i\in[n],T\times n}=\left( \frac{Y_1-[\xi_1,\cdots,\xi_p]\beta_1}{\sigma_1},\ldots,\frac{Y_n-[\xi_1,\cdots,\xi_p]\beta_n}{\sigma_n}\right)_{T\times n}$, and the matrix $A$ is the adjacency matrix, with $l_A^{-1}$ and $u_A^{-1}$ being its corresponding minimal and maximal eigenvalue. For the prior hyper parameters, we choose $s_0=1$, $\nu_0=1e^{-2}$ and $\Lambda_0=1e^{-6}\times I$. Those values are chosen based on the empirical investigation and sensitivity analysis, e.g., $\nu_0=1e^{-2}$ is ``non-informative" enough while still allowing the underlying Dirichlet Process to generate new clusters, and the posterior outputs are quite stable by choosing $\Lambda_0=\lambda_0\times I$ with $\lambda_0$ taking values in the range of $ [1e^{-10},1e^{-2}]$.


\section{Bayesian Inference}\label{sec:bayesian_inference}
In this section, we discuss the posterior sampling method and model selection criterion for the proposed clustering approach.
\subsection{Bayesian Computation}\label{sec:bayes_comput}
To facilitate the posterior sampling of $(\{\beta_c,\sigma^2_c\}_{c\in\mathcal{C}},\mathcal{C},\phi)$, we consider the Gibbs sampling scheme for the following three quantities iteratively: (1) $\{\beta_c,\sigma^2_c\}_{c\in\mathcal{C}}\mid\mathcal{C},\phi,\{Y_i\}_{i=1}^n,\{\xi_{j}\}_{j=1}^p$, (2) $\mathcal{C}\mid\phi,\{\beta_c,\sigma^2_c\}_{c\in\mathcal{C}},\{Y_i\}_{i=1}^n,\{\xi_{j}\}_{j=1}^p$, and (3) $\phi\mid\mathcal{C},\{\beta_c,\sigma^2_c\}_{c\in\mathcal{C}},\{Y_i\}_{i=1}^n,\{\xi_{j}\}_{j=1}^p$. The algorithm is summarized below. 

\begin{algorithm}[htp]\label{algo}
\SetAlgoLined
\kwInit{Initial partition: $\mathcal{C}$, and initial cluster parameters: $\{\beta_c,\sigma^2_c\}_{c\in\mathcal{C}}$.}
\For{$iter=1,2,\cdots M$}{
    Step (1): Update $\{\beta_c,\sigma^2_c\}_{c\in\mathcal{C}}$ conditioning on $\mathcal{C}$ and $\phi$.\\
    \For{$c\in\mathcal{C}$}{
        Sample parameters for cluster $c$ from the full conditional distribution:\\
        $p(\beta_c,\sigma^2_c\mid\{Y_i\}_{i\in c},\{\xi_{j}\}_{j=1}^p,\phi,\mathcal{C})\propto \Pi_{i\in c}f(Y_i\mid\{\xi_{j}\}_{j=1}^p,\beta,\sigma^2,\phi)\pi(\beta,\sigma^2)$\\
        $\propto (\sigma^{-2}_c)^{a_n}\exp\{-\sigma^{-2}_cb_n\}(\sigma^{-2}_c)^{p/2}\exp\{-\frac{1}{2}\sigma^{-2}_c(\beta-\mu_n)^T\Lambda_n(\beta-\mu_n)\}$.
    }
    Step (2): Update $\mathcal{C}$ conditioning on $\{\beta_c,\sigma^2_c\}_{c\in\mathcal{C}}$ and $\phi$.\\
    \For{$i=1,2,\cdots n$}{
        Remove index $i$ from a $c\in\mathcal{C}$, denote resulting partition by $\mathcal{C^*}$.\\
        Put $i$ back into a $c\in\mathcal{C^*}$ with probability $\propto \sum_{j\in c}W(h)_{i,j}\times f(Y_i\mid\{\xi_{j}\}_{j=1}^p,\beta_c,\sigma^2_c,\phi)$,\\
        or create a new cluster for $i$ with probability $\propto \alpha\times\int f(Y_i\mid\{\xi_{j}\}_{j=1}^p,\beta,\sigma^2,\phi)dG_0$.\\
        Let $\mathcal{C}$ be the resulting partition.
    }
    Step (3): Update $\phi$ conditioning on $\{\beta_c,\sigma^2_c\}_{c\in\mathcal{C}}$ and $\mathcal{C}$.\\
    Propose a new value for $\phi=\phi_{iter}$ around $\phi_{iter-1}$ using a normal proposal, truncated between $(\ell_A,u_A)$.
}
where $a_n=\frac{T\times |c|+\nu_0}{2}$, $b_n=\frac{1}{2}(\nu_0s_0^2+\mu_0^T\Lambda_0\mu_0+tr\{[Y_i]_{i\in c,T\times |c|}^T[Y_i]_{i\in c,T\times |c|}(I-\phi A_{c,c})\}-\mu_n^T\Lambda_n\mu_n)$, $\Lambda_n=\mathbf{1}_{|c|}^T(I-\phi A_{c,c})\mathbf{1}_{|c|}\xi^T\xi+\Lambda_0$, $\xi=[\xi_1,\cdots,\xi_p]_{T\times p}$, and $\mu_n=\Lambda_n^{-1}(\xi^T[Y_i]_{i\in c,T\times|c|}(I-\phi A_{c,c})\mathbf{1}_{|c|}+\Lambda_0\mu_0)$. 
 \caption{posterior sampling scheme for gwDP-CAR model.}
\end{algorithm}
In Algorithm 1, each column in $[Y_i]_{i\in c,T\times|c|}$ is the scaled growth rate curve of a specific state assigned to the cluster $c$. In Step (1), we can directly sample from a closed form distribution, since the prior specified in Section~\ref{sec:hier_model} is conjugate to the likelihood function, which results in the multivariate-normal-inverse-gamma posterior. Step (2) requires an enumeration of the index set, i.e., $[n]\equiv\{1,2,\cdots, n\}$, and then updating the allocation of each state via the full conditional distribution of $i\mid[n]\setminus\{i\}$, according to the P\'{o}lya Urn scheme of gwCRP as introduced in Section~\ref{sec:gwCRP}. Step (3) is the most challenging part, as sampling from a closed form is unavailable for $\phi$. We consider the Metropolis-Hastings algorithm with a normal proposal to sample $\phi$. In addition, we show in the supplementary materials that the log-likelihood function of $\phi\mid\mathcal{C},\{\beta_c,\sigma^2_c\}_{c\in\mathcal{C}},\{Y_i\}_{i=1}^n,\{\xi_{j}\}_{j=1}^p$ is proportional to $C_1\times \log(\det(I-\phi A))+C_2+C_3\times\phi$ for some constants $C_1,C_2,C_3$, which is approximately concave within the range $(\ell_A,u_A)$, and hence guaranteeing a quick convergence. We refer our readers to supplementary materials for more details and derivation of the full conditional distribution.

\subsection{Model Selection}\label{sec:model_select}
We select the  spatial smoothness tuning parameter $h$ as defined in \eqref{eq5} based on the Logarithm of the Pseudo-Marginal Likelihood \citep[LPML;][]{geisser1979predictive}, which is defined as
$$\text{LPML}=\sum_{i=1}^n\log(\text{CPO}_i),$$
where $\text{CPO}_i$ is the Conditional Predictive Ordinate statistic for the state $i$, defined as
$$\text{CPO}_i=f(y_i\mid y^{(-i)}),$$
where $y^{(-i)}$ refers to the entire data set excluding the $i$-th state. A Monte Carlo estimate of CPO \citep{chen2012monte} can be obtained as
$$\widehat{\text{CPO}_i}=\left\{\frac{1}{M}\sum_{l=1}^M\frac{1}{f(y_i\mid\theta_l)}\right\}^{-1},$$
where $\theta_l$ is the $l$-th posterior sample from the MCMC. In the context of our problem, $\theta_l$ should be $\{\beta_i,\sigma^2_i\}_{i=1}^n$ obtained in the $l$-th iteration of the MCMC, following the notation given in Section~\ref{sec:hier_model}. This gives the estimated LPML as
$$\widehat{\text{LPML}}=\sum_{i=1}^n\log(\widehat{\text{CPO}_i}).$$
In general, the model with a larger LPML value should be preferred, and therefore we will choose the $h$ that maximizes the LPML.
\section{Simulation}\label{sec:simu}
\subsection{Simulation Setup and Evaluation Metrics}\label{sec:simu_setup}
\begin{figure}[H]
\begin{center}
\centerline{\includegraphics[width=1\textwidth]{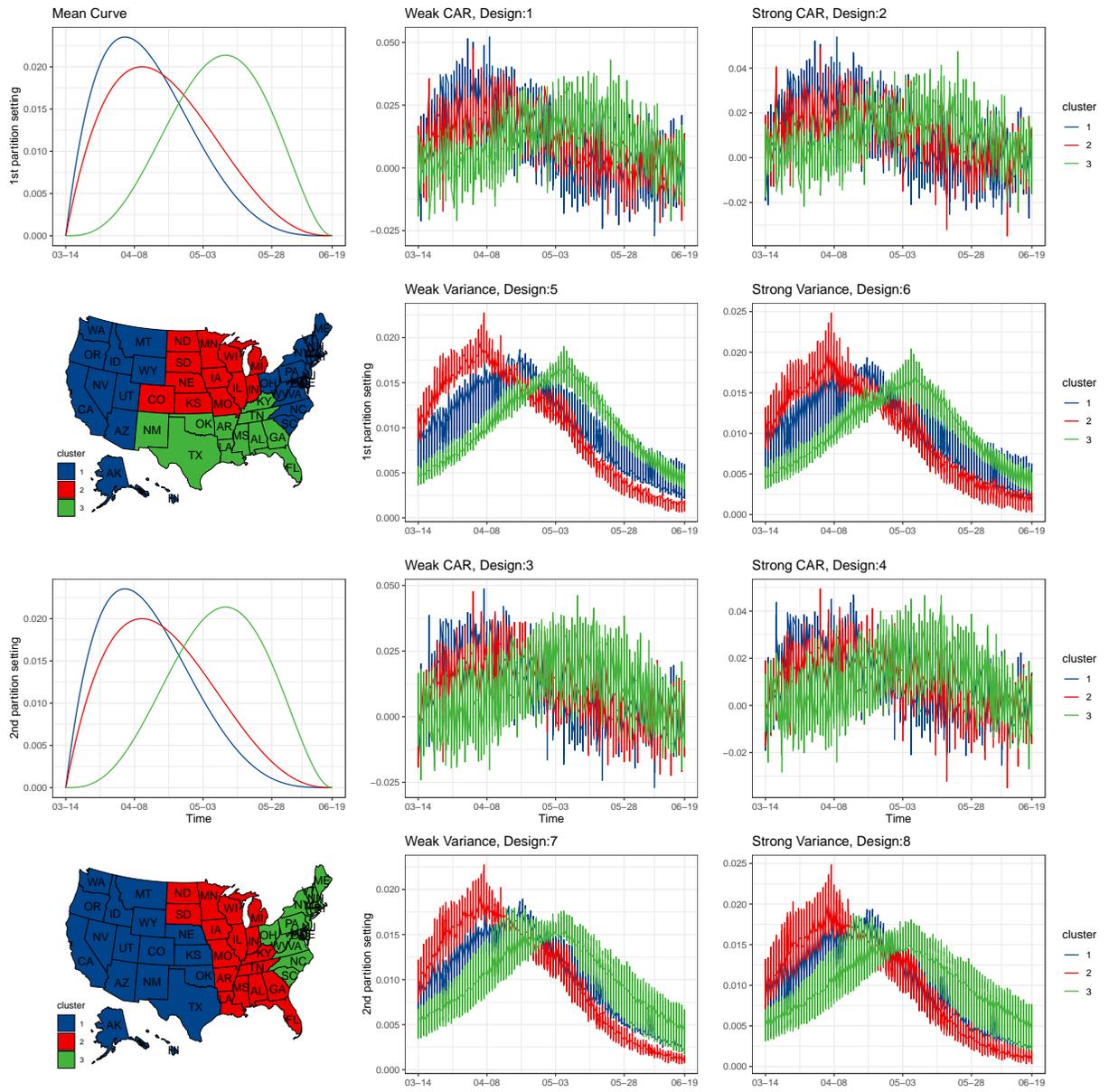}}
\end{center}
\caption{\label{fig:two}Two partition settings and one realization of each design}
\end{figure}

For simulation data generation, we consider two clustering structures, as shown in the mean curve plots at the left side column of Figure~\ref{fig:two}, where in both structures three clusters are considered with two of them (blue and red curves) being close to each other. The main difference is that the blue colored cluster in the first setting contains two spatially in-contiguous blocks, which reflects the spatial patterns that we have observed in the preliminary data analysis. For each clustering structure, we consider two generation schemes. In the first scheme, the data is generated based on the proposal functional model where the cluster-wise scaled growth rate curve is set equal to a normalized Beta density curve to mimic different outbreak timings that we have observed in the real data plus Gaussian random error. More specifically, let $n=51$ and $T=98$, and we consider
\begin{align*}
& f_1(t)=t\cdot (1-t)^{3.5};f_2(t)=t\cdot (1-t)^{2.5};f_3(t)=t^3\cdot (1-t)^{2}, \\
& \mu_j(t)=\frac{f_j(t)}{\sum_{t}f_j(t)}\text{, for $t=i/97$, $i=0,1,\cdots,97$} \\
& Y_k^0(t)=\mu_{c_k}(t)+\sigma\times \epsilon_k\text{, for $k=1,\cdots,51$,} \\
& \text{vec}([\epsilon_{1},\epsilon_{2},\cdots,\epsilon_{51}])\sim \mathcal{N}(0,(I-\phi A)_{51\times 51}^{-1}\otimes I_{98\times 98}). \\
& Y_k(t)=\frac{Y_k^0(t)}{\sum_{t}Y_k^0(t)}.
\end{align*}
We set $\sigma^2=6e^{-5}$ and consider two noise levels within the CAR setting, a weaker noise level with $\phi=0.01$, and a stronger level with $\phi=0.15$. Figure~\ref{fig:two} presents a random draw under two clustering structures and two noise levels, denoted by, design 1 to 4.  

For the second data generating scheme, we consider a SIR model as follows, 
$$\frac{dS}{dt}=-\frac{\beta IS}{N},\ \frac{dI}{dt}=\frac{\beta IS}{N}-\gamma I,\ \frac{dR}{dt}=\gamma I,$$
where $S$ is the susceptible population, $I$ is the infected population, $R$ is the recovery population, $N=S+I+R$ is the total population, and $\beta$ and $\gamma$ are the rates of infection and recovery, respectively. The scaled growth rate curve $Y_i(t)$ for state $i$ on day $t$ is defined as
$$s_i(t)=I_i(t)+R_i(t)-I_i(t-1)-R_i(t-1),\ Y_i(t)=\frac{s_i(t)}{\sum_t s_i(t)}.$$
In order to mimic the scaled growth rate trend observed in the real data, we consider, for each cluster, an SIR model with four turning points in $\gamma$, representing that the state governments gradually have pandemic under control, i.e., 
$$\beta_i=\beta_{c_i},\ \gamma_{i}=\sum_{j=1}^4 \delta_{j,i} I(t_{j-1,c_i}\leq t< t_{j,c_i}).$$
We set the initial value $I_i(0)=5000$, $R_i(0)=500$ and $N$ to be a constant that equals to the population of each state in 2020. A detailed list of parameter values used in our simulation is given in Table~\ref{tab:one}. Under this setting, the scaled growth rate curves are expected to increase in the first two phases and decrease in the rest two phases. We also consider two noise levels, with a stronger level ($\text{sd}=0.015$), and a weaker level ($\text{sd}=0.01$). Figure~\ref{fig:two} presents a random sample under each setting (two clustering structure and two noise levels for SIR model), named as design 5 to 8.  



\begin{table}
\caption{Simulation setting and parameter values for the SIR data generation. Here U(a) means a random number generated uniformly from $(a-\sigma,a+\sigma)$ with $\sigma$ being a pre-chosen standard deviation.}
\label{tab:one}
\begin{center}
\begin{tabular}{lllllllllll}
\hline
  $c_i$ & $\beta_{c_i}$ & $t_{0,c_i}$ & $t_{1,c_i}$ & $t_{2,c_i}$ & $t_{3,c_i}$ & $t_{4,c_i}$& $\delta_{1,c_i}$ &$\delta_{2,c_i}$ & $\delta_{3,c_i}$ & $\delta_{4,c_i}$ \\
\hline
1 & 0.08& 03/14 & 04/03 & 04/23 & 05/13 & 06/20 & U(0.05) & U(0.065) & U(0.095) & U(0.11)\\
2 & 0.11& 03/14 & 03/24 & 04/08 & 05/08 & 06/20 & U(0.06) & U(0.095) & U(0.125) & U(0.16)\\
3 & 0.14& 03/14 & 04/18 & 05/08 & 05/18 & 06/20 & U(0.11) & U(0.125) & U(0.155) & U(0.17)\\
\hline
\end{tabular}
\end{center}
\end{table}

To evaluate the clustering performance, we adopt the widely-used rand index \citep{rand1971objective} that compares the clustering result with the ground truth. More specifically, for two partitions $\mathcal{C}_1$ and $\mathcal{C}_2$ implemented on $n$ observations, rand index is defined as
RI$=\frac{a+b}{{n\choose 2}}$, where $a$ denotes the number of observed pairs that are in the same cluster in $\mathcal{C}_1$ and $\mathcal{C}_2$ simultaneously, called by ``true positive", while $b$ denotes the number of observation pairs that are in different clusters in $\mathcal{C}_1$ and $\mathcal{C}_2$ simultaneously, called by ``true negative". Rand index takes values from 0 to 1, with a larger value indicating a higher level of coherence. 

\subsection{Simulation Results}\label{sec:simuresult}
We compare our proposed method with two competing methods. The first one is the $K$-means method for functional data, which is implemented using the R package \textit{kml} \citep{genolini2016package}. The second method is the funcitonal EM method implemented using the R package \textit{funFEM} \citep{bouveyron2015funfem}. In order to conduct a fair comparison, we use the Calinski Harabasz Criterion \citep{calinski1974dendrite} in the $K$-means method and the Bayesian Information Criterion \citep[BIC,][]{schwarz1978estimating} in the EM method for choosing the number of clusters $k$ within a candidate pool $k \in \{2,3,\ldots,5\}$. For our method, we select $h$ based on the highest LPML value, and run MCMC for 4000 iterations with the first 2000 iterations as burn-in. The posterior samples are summarized by Dahl's method \citep{dahl2006model} to obtain the estimated clustering results. 

We summarize the average RI values based on 100 Monte Carlo replicates for 8 simulation designs in Table~\ref{tab:two}. We find that the proposed gwDP has a clear advantage over DP under all simulation designs, which confirms the benefit of incorporating spatial heterogeneity, i.e., gwDP can utilize the geographical information to refine weights when sampling from CRP and hence avoid the common issue for conventional DP where the posterior sample are trapped in a local optima. When the data is generated from the proposed model as in Designs 1-4, both DP and gwDP have a significantly better performance than those of the $K$-means and the EM. This advantage becomes less apparent in Designs 5-8 (gwDP still gives the highest RI value 3 out of 4 designs) when the data is generated from a SIR scheme. This can be explained by the empirical observation (see, Figure \ref{fig:two}) that the scaled growth rate curves generated from the SIR scheme (Designs 5-8) are smoother compared to those generated from our model specification  (Designs 1-4), which indicates that the state-wise residuals on different time point may no longer be independent. The violation of independence assumption for our model hence contribute to the negative effect on the clustering accuracy. In conclusion, our model is considerably powerful in the independent case, and is still very competitive even if the residuals are dependent.

\begin{figure}[htp]
\minipage{0.5\textwidth}
  \includegraphics[width=\linewidth]{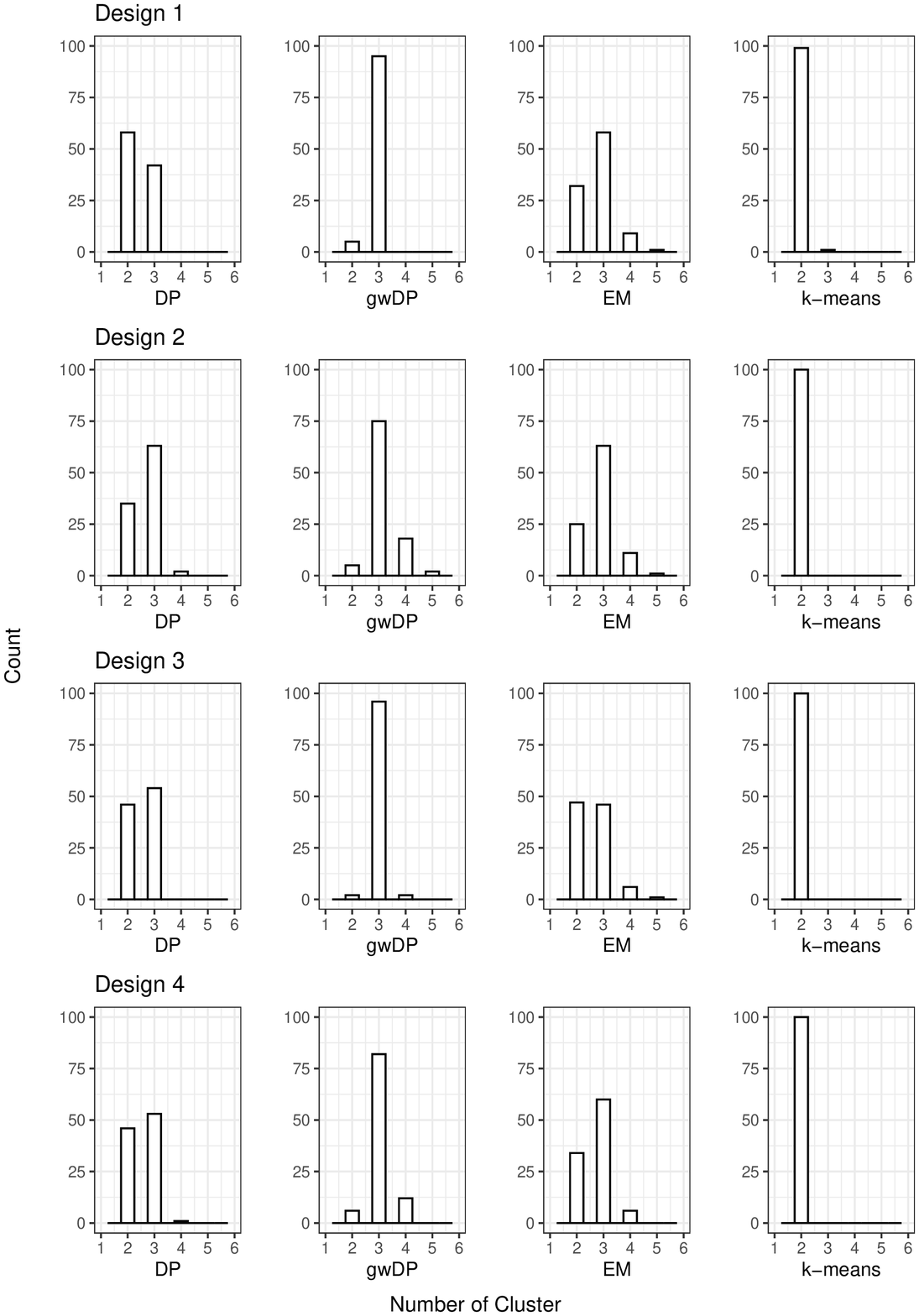}
\endminipage\hfill
\minipage{0.5\textwidth}
  \includegraphics[width=\linewidth]{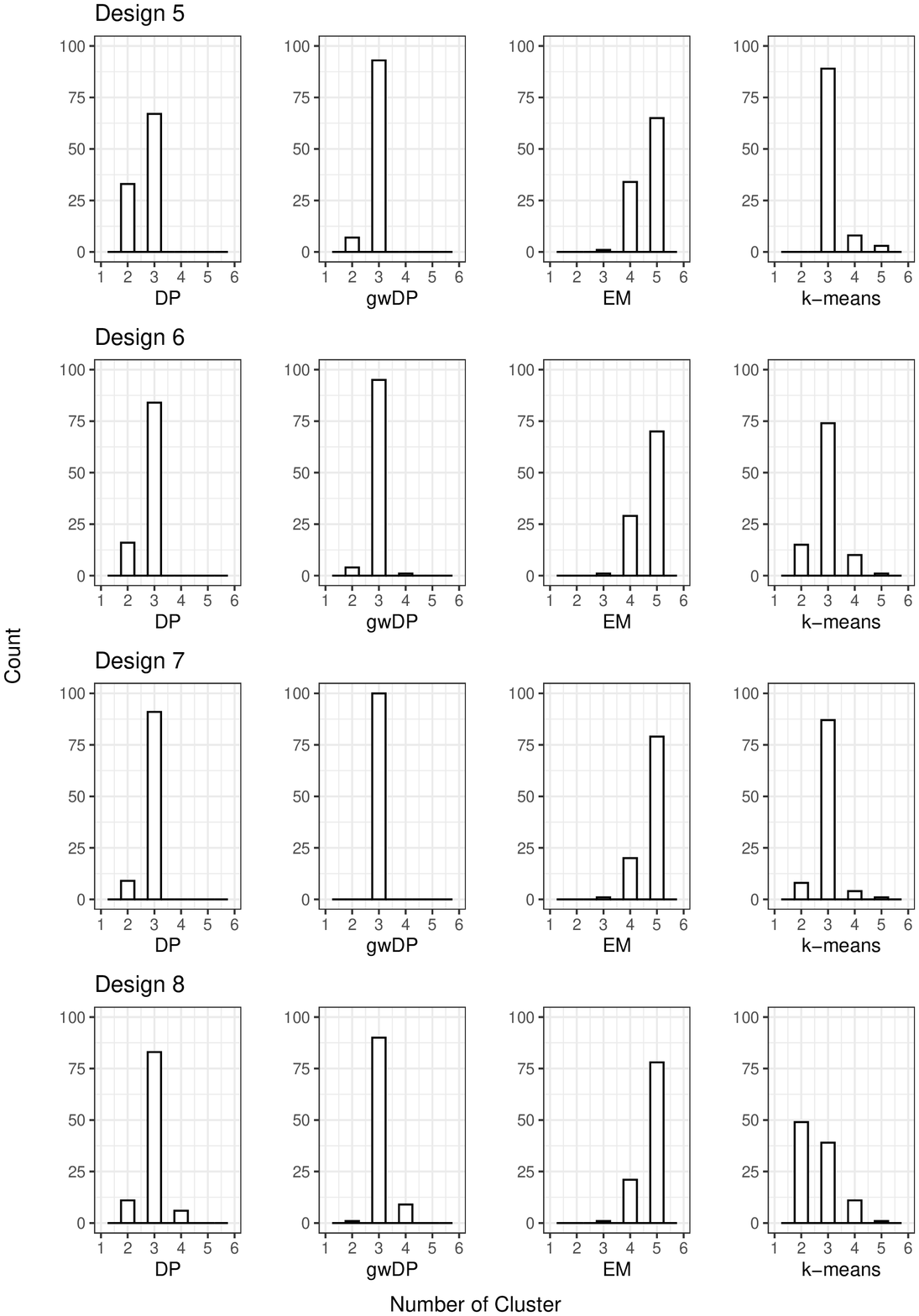}
\endminipage\hfill
\caption {\label{fig:three}Comparison of inference on $k$ among four methods}
\end{figure}

Next we present the histograms for the selected $k$ in Figure~\ref{fig:three}. It is clear that the proposed gwDP has an excellent performance in terms of choosing the correct number of clusters $(k=3)$ under all designs. The conventional DP is likely to underestimate $k$, because the high variance would conceal the difference between clusters, while the gwDP can unveil such discrepancy by utilizing the geographical information. The $K$-means method, a distance-based method, works relatively well if the variance is not horribly large (the last four designs), but fails in the large variance case (the first four designs) for the same reason with the conventional DP's underestimation of $k$. The EM method, fails to provide a valid estimation on $k$ in the last four designs. Overclustering happens since the EM method that neglects geographical facts would prefer more clusters, even if the size for some clusters is abnormally small. These small clusters usually contain two or three curves, which are generated far away from their cluster average by chance. 

\begin{table}[htp]
    \centering
        \caption{Average RI over 100 simulation replicates the proposed gwDP and three competing methods}
    \label{tab:two}
   \begin{tabular}{lcccccc}
    \toprule 
Design & Structure & $\phi$ (1st Scheme) & gwDP & DP & $K$-means & EM\\
    \midrule 
1& 1st & 0.01 & \textbf{0.948} & 0.817 & 0.737 & 0.708\\
2& 1st & 0.15 & \textbf{0.902} & 0.842 & 0.736 & 0.714\\
3& 2nd & 0.01 & \textbf{0.966} & 0.862 & 0.773 & 0.772\\
4& 2nd & 0.15 & \textbf{0.925} & 0.849 & 0.773 & 0.781\\
    \midrule
& & sd (2nd Scheme) & & &  &\\    
    \midrule
5& 1st & 0.010 & \textbf{0.941} & 0.865 & 0.931 & 0.833\\
6& 1st & 0.015  & \textbf{0.802} & 0.768 & 0.778 & 0.756\\
7& 2nd & 0.010  & \textbf{0.876} & 0.853 & 0.865 & 0.837\\
8& 2nd & 0.015  & 0.781 & 0.764 & \textbf{0.785} & 0.753\\
        \bottomrule
    \end{tabular}
\end{table}

\section{COVID-19 Data Analysis}\label{sec:real_data}
In this section, we apply the proposed approach to study the COVID-19 scaled growth rate curves discussed in Section \ref{sec:data}. Results from a state level analysis and a New York county level analysis will be presented. We run 16000 MCMC iterations with the first 8000 iterations as burn-in. The hyper-parameters are set as $\nu_0=1e^{-2}$ and $\Lambda_0=1e^{-6}I$. We choose $h=0.511$ for state level data and $h=1.564$ for county level data, for which both values are obtained by maximizing LPML. 

\subsection{State Level Analysis}\label{sec:state_level}
\begin{figure}[htp]
\minipage{0.5\textwidth}
  \includegraphics[width=\linewidth]{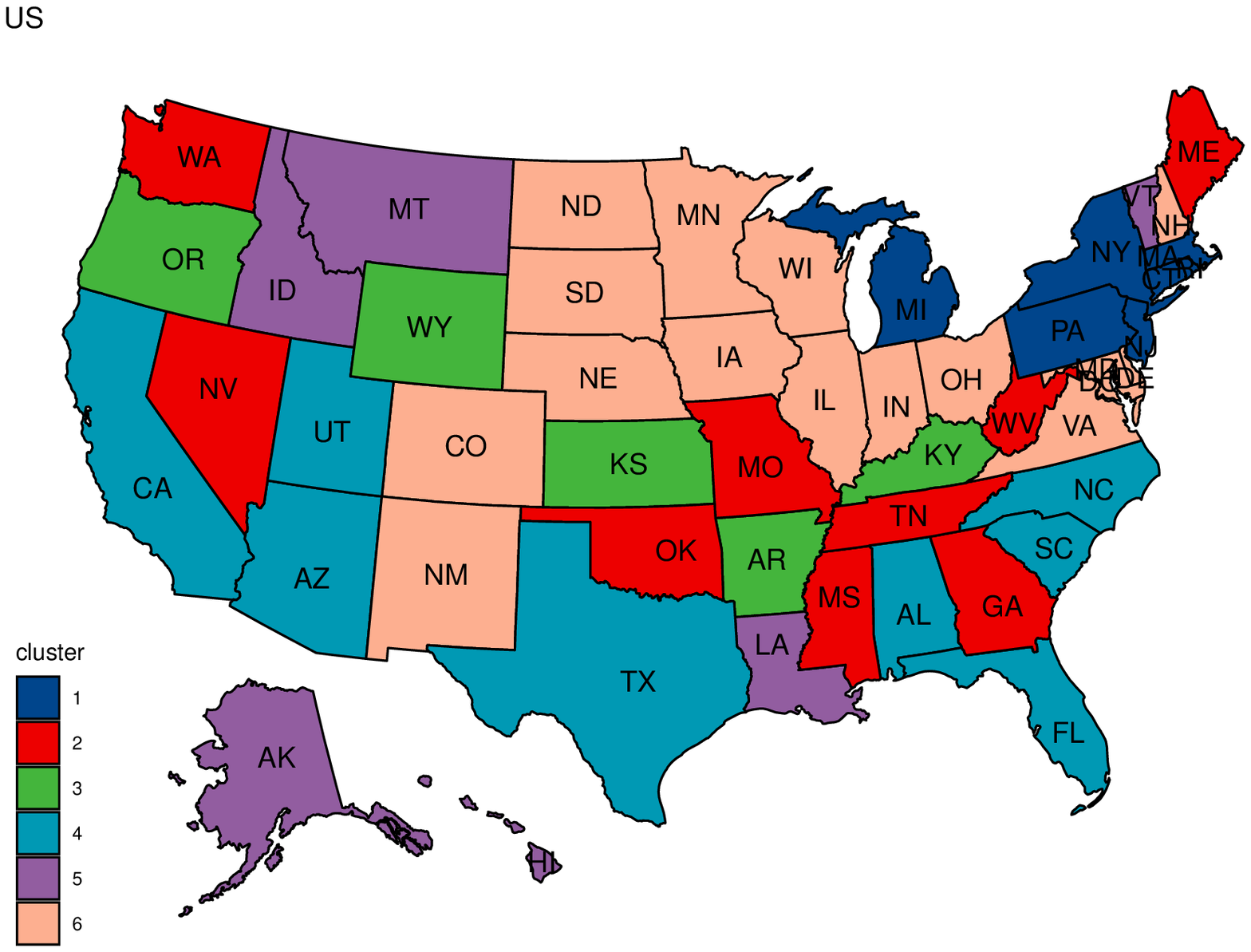}
\endminipage\hfill
\minipage{0.5\textwidth}
  \includegraphics[width=\linewidth]{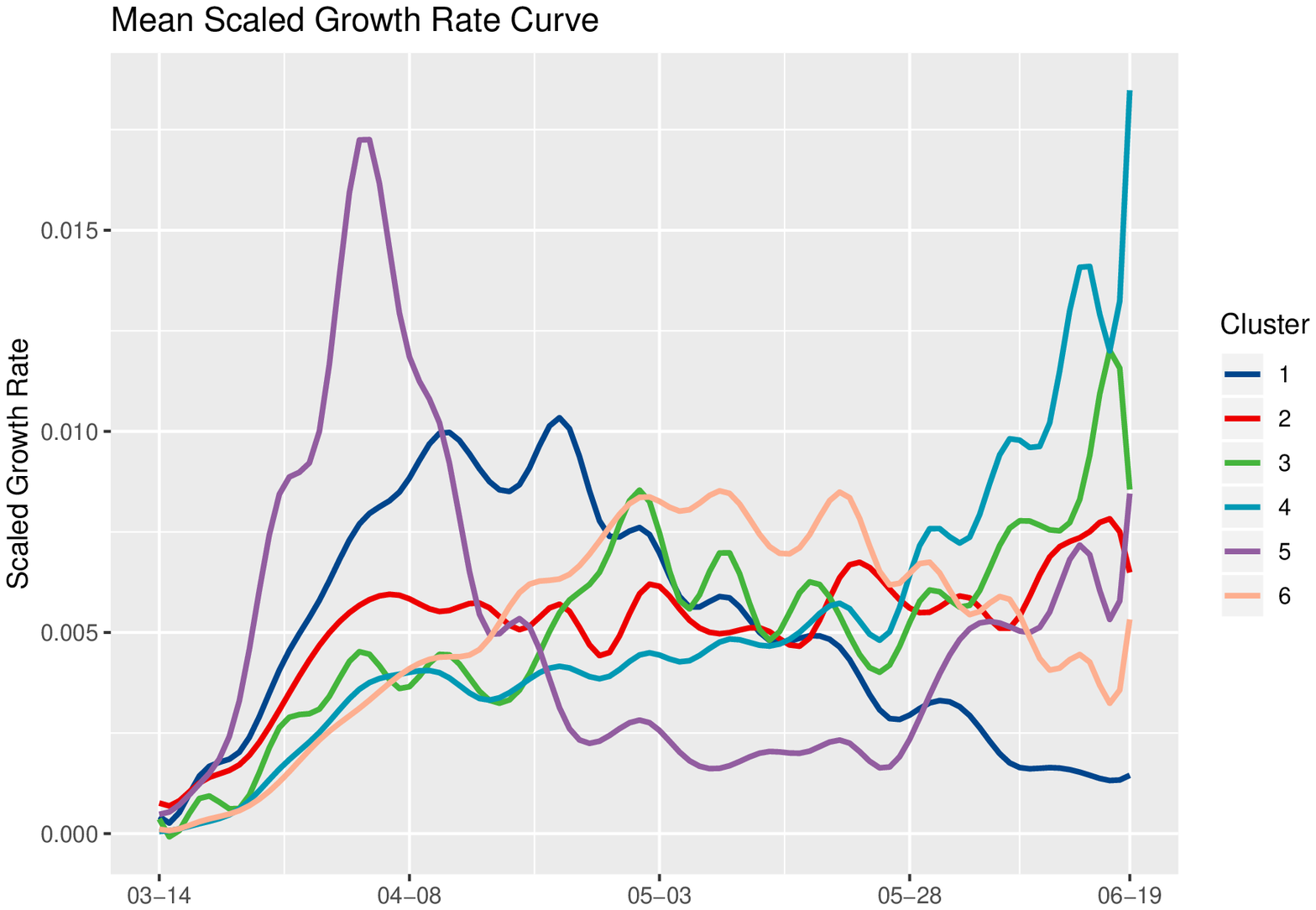}
\endminipage\hfill
\newline
\minipage{0.5\textwidth}
  \includegraphics[width=\linewidth]{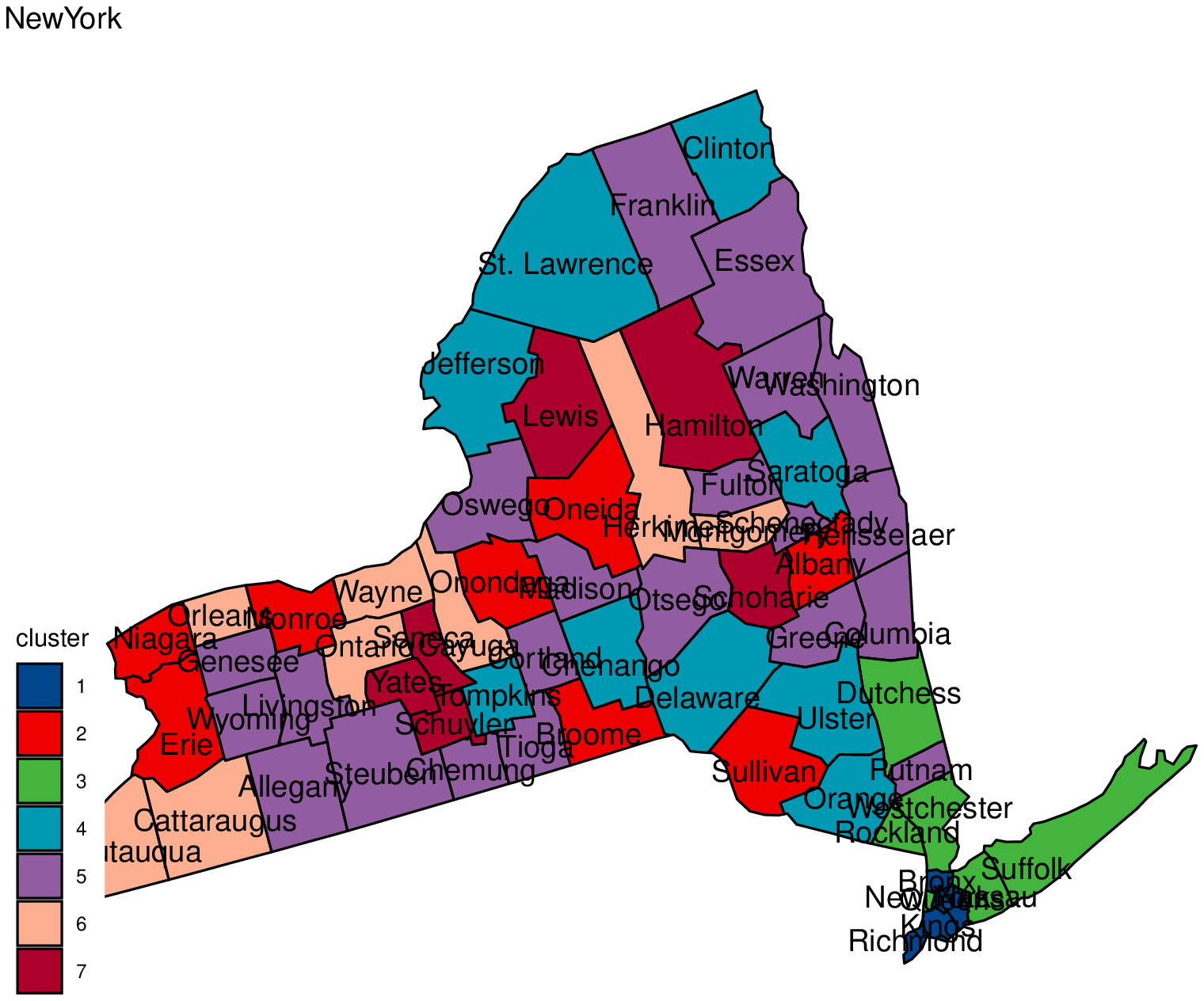}
\endminipage\hfill
\minipage{0.5\textwidth}
  \includegraphics[width=\linewidth]{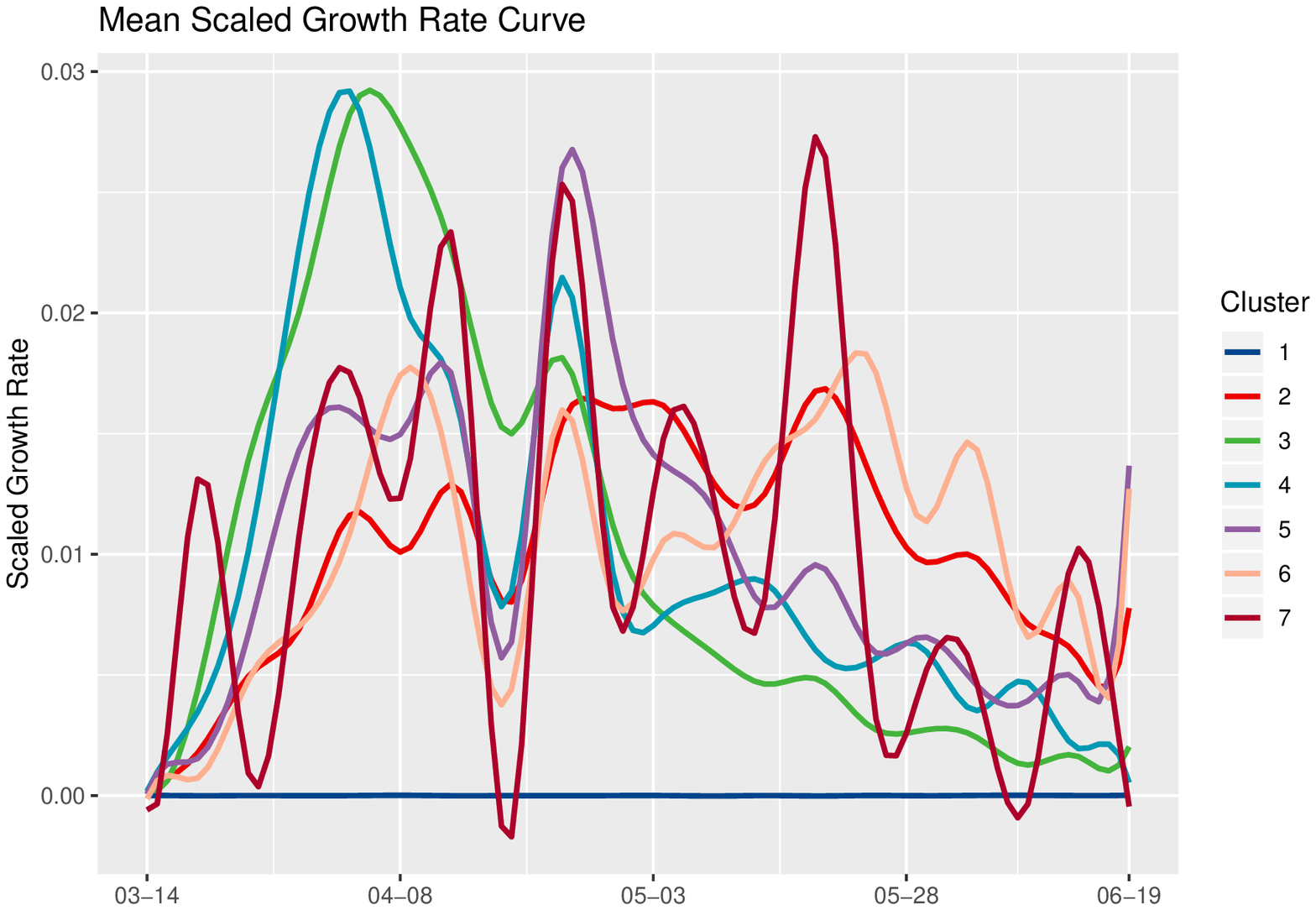}
\endminipage\hfill
\caption{\label{fig:four}Left: State-level cluster assignment; Right: Mean scaled growth rate curve}
\end{figure}

We summarize the state level clustering analysis results in Figure~\ref{fig:four}, including the cluster assignment for each state and the \textit{mean scaled growth rate curve}, whose coefficients of basis function are obtained by averaging the $\beta_{c_i}$ over all iterations and replications where the cluster assignment is identical to the reported one. For the selected $h$, we re-run the model for another 100 replications with different random seeds to evaluate the stability of the reported cluster assignment results. We find that the average Rand Index obtained from the 100 replications relative to the reported clustering results is 0.909, which implies a high level of concordance and hence confirms the desired stability property of our clustering results. Besides, the $95\%$ credible interval for $\phi$ is $[0.020,0.057]$, which indicates a mild level of spatial correlation, as the upper bound of $\phi$ is 0.184.
As shown on the left side of Figure~\ref{fig:four}, the state-level growth rate curves are clustered into six groups. There are three main patterns in the overall trend of the mean scaled growth rate curves: (1) Rapid increasing and decreasing trend, with a rise in late May, e.g., cluster 5. (2) Slowly increasing and decreasing trend, e.g., cluster 1 and 6. (3) Constantly increasing trend, e.g., cluster 2, 3 and 4. In the rest of this section, we provide more detailed discussion for each trend and also study those states with high total confirmed cases.

\textbf{Pattern 1} The representative states are Montana and Idaho in cluster 5. Inside this pattern group, the mean scaled growth rate shows a spike in late March, followed by a rapid decay, which indicates COVID-19 was well controlled by these states before late May. One possible explanation is that, these states were not reopened by their governors until early June. Some states, such as Idaho and Vermont, though underwent reopening in middle April, only allowed essential businesses to reopen. Besides, there was an upward trend for both states around June 4, consistent with the reopening policies issued by these states in early June.

\textbf{Pattern 2} Though classified as having the same pattern, the discrepancy between the mean scaled growth rate curves of these two clusters (1 and 6) is considerably large. Each cluster involves a certain outbreak timing of COVID-19: (1) The cluster 1, consisting of New York, New Jersey, and Michigan, which are known as the states where the pandemic began in the United States, achieved its peak of daily new confirmation in middle April. Some negative events, such as in Massachusetts, were reported that the federal government impounded a shipment of three million masks on March 18, and some major hospitals had to reuse masks due to lack of medical supply. These findings provided evidence to explain why the states in cluster 2, unlike the states in the pattern group 1, missed the opportunity of controlling pandemic spread at its beginning stage. (2) The cluster 6, including  Illinois and Indiana, has a peak around early May. Geographically close to the cluster 1, these states showed a several-day delay in the outbreak timing probably because of the logistics and disease transmissions. Though the state governments showed a quick response to the emergent situation, the curves seemed to suggest a lost of the control for the spread of pandemic at its early stage until mid-May.  

\textbf{Pattern 3} Many states assigned to this group are ranked in Tier 1 in terms of the total number of confirmed cases, such as California, Texas, Florida, and Georgia. All of these states were reopened before the pandemic was under well control. For example, California was reopened in May. Texas, undergoing some gathering events such as strikes, was reopened in middle May. The Florida governor reopened some beaches in middle April; and for Georgia, though, did not undergo a spread out of pandemic at the early stage, lost control for the pandemic because of two reopening policies issued on April 2 and April 24, resulting in rises in daily new confirmation. The third cluster, though assigned to this pattern group, cannot be directly interpreted, as the total number of confirmation for each of these five states is relatively low compared to the other states in this pattern group, which makes the scaled growth rate curve abnormally fluctuated. It is difficult to tell if the trend observed from the mean scaled growth rate curve comes essentially from intrinsic mechanism or by coincidence.  

The cluster assignment of Louisiana seems to be counterintuitive. Though surrounded by many states in the pattern group 3, it is still believed by our model to be in the pattern group 1. Louisiana governors ordered the closure of schools, bars, and casino gaming in middle March, and unlike Texas and Florida, which are spatially close to Louisiana, did not reopen the state until early June. These policies may be the deciding factor to help Louisiana effectively control the pandemic at an early stage. 

Besides the aforementioned findings, we also observe that those states that are geographically contiguous are more likely to be assigned to the same cluster, e.g., the cluster 1 and 6. Meanwhile, this does not exclude the possibility that distant states can still belong to the same cluster. For example, Louisiana and Vermont are assigned to the cluster 5 together with Idaho and Montana, though they are far away from the latter two states. These findings confirm the flexibility of the proposed gwDP method in terms of clustering geographically contiguous and distant states.

\subsection{County Level Analysis}
Next we discuss a county-level analysis for the New York state. Similarly with the state-level analysis, we have conducted 100 replications based on the selected $h$, and found that the rand index of these replications relative to the reported cluster assignment is 0.941 on average, which confirms the stability of the reported cluster assignment results. Besides, the $95\%$ credible interval for $\phi$ is  $[0.108,0.128]$, which indicates a relatively strong spatial correlation, as the upper bound of $\phi$ is 0.180. In the bottom part of Figure \ref{fig:four}, we find that cluster 3 and 4 are of main interest compared to the other clusters, since the total confirmation of the rest clusters is negligible relative to these two clusters. As discussed before, a small total confirmation tend to yield a more fluctuated scaled growth rate curve, which further makes the interpretation of the mean scaled growth rate curve less reliable.  

The growth curves from cluster 3 and 4 look quite similar to the mean scaled growth rate curve of the entire New York state, i.e., a rising trend before April-8th and a decreasing trend afterwards. It makes great sense since the total confirmation of these two clusters accounts for up to 93$\%$ total confirmation of the New York state. Besides, geographically, most nearby counties of New York city are assigned to these two clusters, which agrees with the fact that the contiguous counties 
are more likely to have a similar pattern in their scaled growth rate curves as a result of the pandemic spread. Some assignments could be problematic, such as Clinton and Jefferson, which are the enclaves of cluster 4. The total confirmation of these two counties takes trivial amount (smaller than 0.05$\%$) of New York state's total confirmation, which makes the clustering assignment of these two counties less reliable.

\section{Discussion}\label{sec:discussion}
In this paper, we proposed a new nonparametric Bayesian clustering method for analyzing spatially correlated functional data. Compared to the classical DP model, the proposed method managed to fully utilize the geographical information and had a significantly improved clsutering performance. An computationally efficient MCMC algorithm was also introduced to infer the posterior distributions of both the number of clusters and the clustering configuration. The applications to COVID-19 data resulted in several inspiring conclusions that unveiled the process of the pandemic spread and the investigation of common/different patterns among clusters had led to the discovery of several useful factors related to the pandemic development such as the reopening policy. These findings are also useful for improving the individual state/county level growth rate prediction after taking account for the clusterwise spatial heterogeneity and public health decision making in the future, e.g., to prepare for the next outbreak of COVID-19 or other similar diseases, policy makers may refer to the policies executed by the states assigned to the clusters that have achieved success in controlling the previous spread of COVID-19.  

A few topics beyond the scope of this paper are worth further investigation. First, using multivariate outcome functional data models \citep{kang2014bayesian,cao2019modeling} to incorporate auxiliary information such as demographic information will help improve the clustering and disease prediction accuracy \citep{xue2018bayesian}. It will also be of interest to perform other functional data dimension reduction methods \citep{zhang2018functional} instead of FPCA to extract useful information for clustering purpose. In addition, proposing an efficient sampling algorithm without tuning parameter selection and considering a non-stationary spatial structure are both important future directions. One may also consider a more flexible covariance structure such as the auto-regressive structure in the model. 

\bibliographystyle{chicago}
\bibliography{paper.bib}
\end{document}